\documentclass[a4paper]{jpconf}
\usepackage{graphicx}
\usepackage{amsmath}
\DeclareUnicodeCharacter{2212}{-}

\begin{document}
\title{Tracing Water Masers at their Smallest Scale with VLBI}
\def\cpar{\par \vspace{1.3mm}}
\def\kms{km s$^{-1}$}
\def\vlsr{$V_\text{LSR}$}
\author{J M Vorster$^{1,*}$, J O Chibueze$^{1,2}$, T Hirota$^{3,4}$, G C MacLeod$^{5,6}$}

\address{$^1$Centre for Space Research, Potchefstroom campus, North-West University,
Potchefstroom 2531, South Africa}
\address{$^2$Department of Physics and Astronomy, Faculty of Physical Sciences,  University of Nigeria, \\Carver Building, 1 University Road,
Nsukka 410001, Nigeria}
\address{$^3$National Astronomical Observatory of Japan,
National Institutes of Natural Sciences,
2-21-1 Osawa, Mitaka, Tokyo 181-8588, Japan}
\address{$^4$Department of
Astronomical Sciences, SOKENDAI (The Graduate University for Advanced
Studies), Osawa 2-21-1, Mitaka-shi, Tokyo 181-8588, Japan}
\address{$^5$The Open University of Tanzania, P.O. Box 23409, Dar-Es-Salaam, Tanzania}
\address{$^6$Hartebeesthoek Radio Astronomy Observatory, PO Box 443, Krugersdorp, 1741, South Africa}
\ead{jobvorster8@gmail.com}

\begin{abstract}
The high-mass star-forming region NGC6334I-MM1 underwent an energetic accretion event in January 2015. We report the large-scale ($10 - 100$ AU) and small-scale ($\sim 1$ AU) changes in spatial and velocity structures of 22 GHz water masers as observed with VERA before and during the accretion burst. The masers in the northern bow-shock CM2-W2 brightened, and better traced a bow structure during the burst. In the southern regions, there was both activation and disappearance of associations before and during the burst. We measured the amplitudes, central velocities and FWHMs of about 20 features in each epoch. We found that the linear scale of the brightest feature in CM2-W2 grew from 0.6 AU before the burst to 1.4 AU after the burst, possibly indicating that a larger volume of gas was able to sustain masing action as a consequence of the accretion burst. This feature also had a rapid (0.2 yr) brightness increase by a factor of four, which has been previously reported in long-term single-dish monitoring. We propose that the water maser flare could be explained by an increase of the collisional pump rate due to radiative heating of H$_2$ by increased high energy radiation (UV or X-ray) from the inner protostellar core. We also describe the spot and spectral method of maser proper motion calculations. We argue that for high spectral resolution observations the spectral method is more robust for calculating proper motions than the spot method.
\end{abstract}

\section{Introduction}
The 22 GHz water maser transition is an important tracer of star-forming regions and evolved stars. These masers are known to be highly variable and to sustain population inversion in high-density turbulent or shocked environments \cite{2013ApJ...773...70H}. Individual water maser cloudlets have been observed in W49N with typical linear sizes of 1 AU, and Full Width at Half Maximums (FWHMs) of 0.5 \kms{} to 3 \kms{} \cite{1994ApJ...429..253G}. The association between shocks and 22 GHz water masers makes them good tracers of high-velocity protostellar jets.
\cpar{}
An important application of water masers is proper motion and parallax measurements through multi-epoch Very Long Baseline Interferometry (VLBI) observations \cite{2017MNRAS.467.2367B,2020PASJ...72...50V}. The variability of water masers can make it difficult to identify persistent water masers which trace the same gas. Two assumptions are commonly made in water maser proper motion observations. Firstly, that the proper motions of water masers trace gas motions and not just the propagation of pumping conditions \cite{2006A&A...447L...9G}. Secondly, that there is little or no acceleration in the water masers \cite{2010A&A...517A..71S}. The second assumption is equivalent to assuming a near constant radial velocity, \vlsr{}, for the maser features, and linear displacement over time in right ascension and declination. The time dependant environment in the high-mass star-forming region NGC6334I-MM1 due to the recent accretion burst \cite{2018MNRAS.478.1077M,2018ApJ...866...87B} might serve as a novel physical environment in which to test these assumptions. 
\cpar{}
NGC6334I is a high-mass star-forming region which contains nine-millimetre cores and is at a distance of 1.3 kpc \cite{2016ApJ...832..187B}. In January 2015, the largest millimetre core, MM1, underwent flaring in multiple maser species and had an increase in  luminosity by a factor of 16.3$\pm$4.4 \cite{2018MNRAS.478.1077M,2021ApJ...912L..17H}. The region hosts multiple outflows, with bright water masers at the edges of an NW-SE bipolar outflow \cite{2018ApJ...866...87B,2021ApJ...908..175C}. There is still uncertainty about the exact mechanism that caused the flaring of water masers $\sim 2700$ AU away from the accretion bursting source, given that the masers of this transition are understood to be collisionally pumped \cite{2021saip.confE...1V}. Looking closely at individual maser features before and after the accretion burst might shed some light on the physical mechanisms at work in the water maser flare.
\cpar{}
The aim of this experiment is to study the spatial and velocity structure of water masers in NGC6334I at a high angular resolution with VERA before and during the accretion burst.
\section{Observations and Data Reduction}
We did seven epochs of observations of 22 GHz water masers with the VERA VLBI array \cite{2020PASJ...72...50V}. The observations were on epochs 2014.7, 2014.9, 2015.1, 2015.3, 2015.9, 2016.1 and 2016.2. The phase tracking centre was ($\alpha, \delta)=(17^\text{h}20^\text{m}53.377^\text{s},
−35^\circ 46'55.808''$, J2000). The channel width of our observations was 0.44 \kms{}. The correlated visibilities were calibrated and imaged with the Astronomical Image Processing System (AIPS), with the same procedure than is described in \cite{2021ApJ...908..175C}. The synthesized beam size was 1.3 $\times$ 3.3 milliarcseconds. Self-calibration was used on a bright velocity channel to improve the dynamic range of the images. After the data was imaged, the SAD task was used to fit 2D Gaussians to each channel, to get the positions and radial velocities of the maser spots. Spatial precision smaller than the beam width can be obtained for bright maser emission, as spatial precision in VLBI is inversely proportional to the signal-to-noise ratio of a source \cite{2006A&A...452.1099P}. In this work a maser ``spot" refers to a single 2D Gaussian fit in a single channel map, while a maser ``feature" refers to a collection of maser spots, with a velocity gradient in space and a Gaussian spectral distribution. It has been argued that a feature represents a single physical maser cloudlet \cite{1994ApJ...429..253G}.
\section{Results}
\begin{figure}
    \centering
    \includegraphics[width = \textwidth]{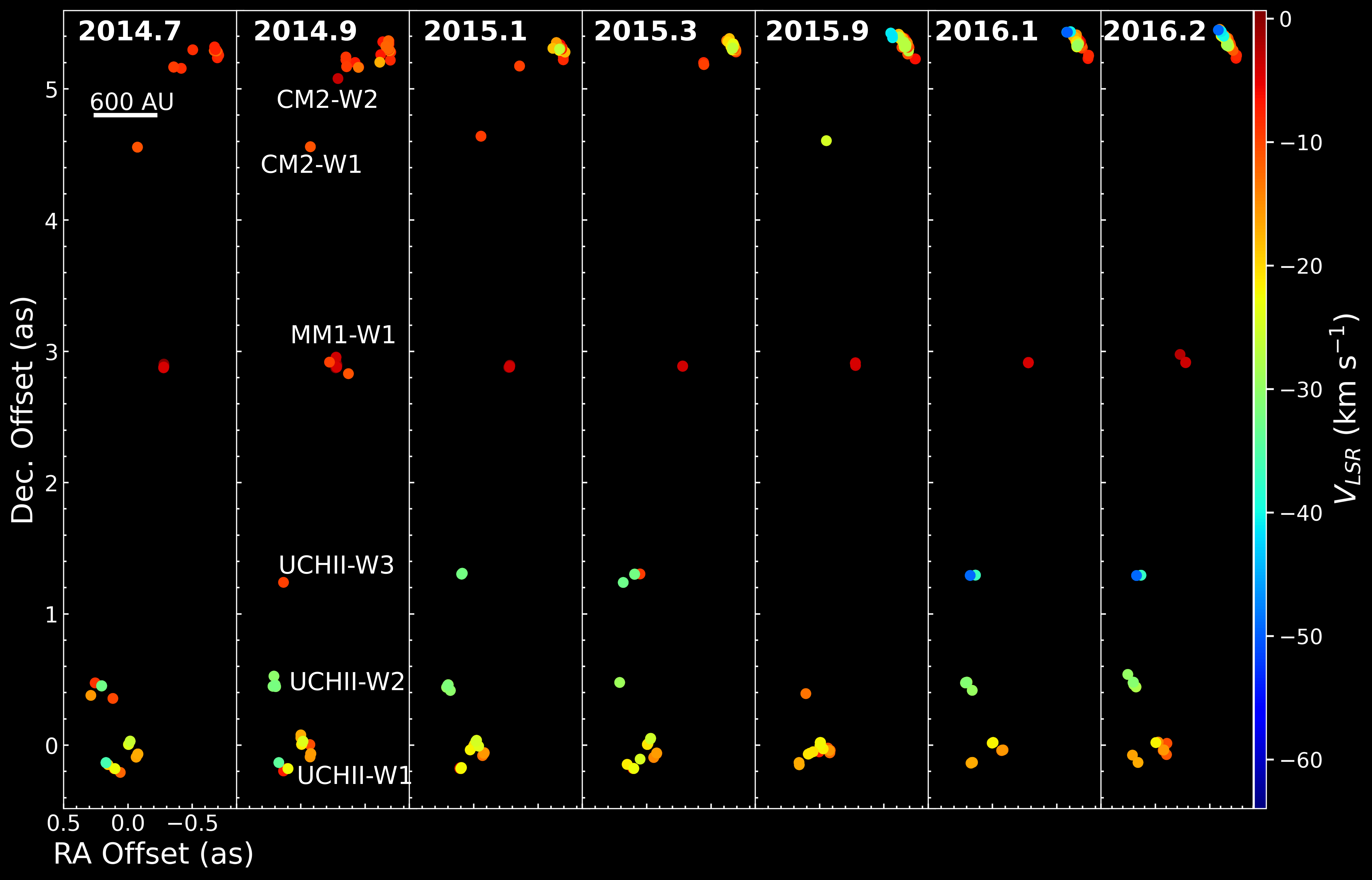}
    \caption{Positions of the water masers as detected by VERA for each epoch. The position of the spots indicates the position of the maser spots while the colour indicates the radial velocity according to the colour scale shown at the right of the figure. The text in the upper left of each panel indicates the date of the observation. The white reference line shows the linear distance scale assuming $d =$ 1.3 kpc. The names of each region are shown in white text in the second panel.}
    \label{fig:VERA_spotmaps}
\end{figure}

\begin{figure}
    \centering
    \includegraphics[width=\textwidth]{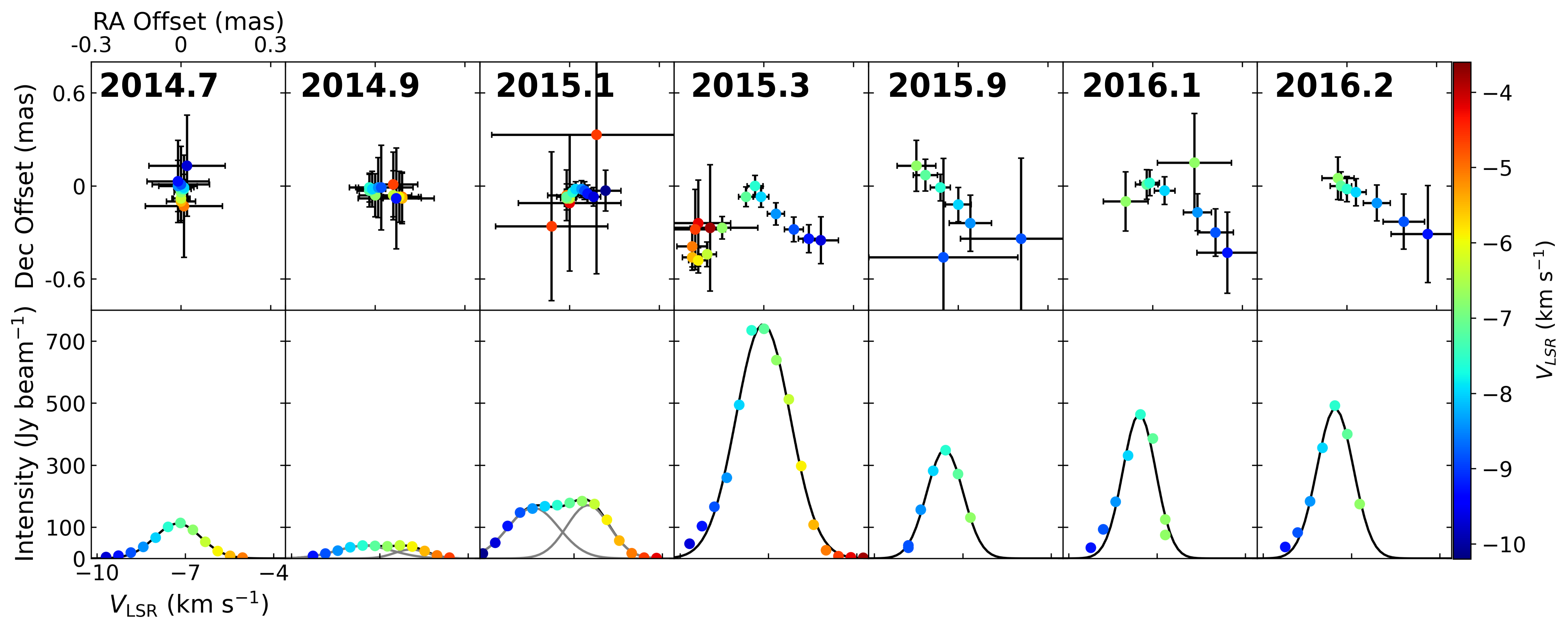}
    \caption{Top: Positions and radial velocities of spots in the brightest maser feature in CM2-W2 over time. The offsets are in terms of the centre position of the feature. The scale is the same for each panel. Bottom: The spectral profile of the feature for each epoch. The black lines show the Gaussian fit to the maser spots. The colour of the spots in each panel indicates the radial velocity according to the colour bar on the right-hand side.}
    \label{fig:reference_feature}
\end{figure}

We detected water masers in CM2-W1, CM2-W2, MM1-W1, UCHII-W1, UCHII-W2 and UCHII-W3 according to the nomenclature of \cite{2018ApJ...866...87B}. Figure \ref{fig:VERA_spotmaps} shows the positions and radial velocities of the maser spots for each epoch. Note that some of the spots in Figure \ref{fig:VERA_spotmaps} are likely spurious detections from imaging artefacts. Imaging artefacts can be identified as one or two faint isolated maser spots with a symmetric spatial distribution around brighter spots.
\cpar{}
We identified features that were a cluster of spatially close ($\sim 1$ AU) maser spots which have a Gaussian spectral distribution. The requirement for maser spots in a feature to have Gaussian spectral distributions filter out spurious detections. We fit Gaussian functions to the maser features' spectral distributions. We detected 15, 23, 28, 27, 22, 21 and 22 features in the seven epochs respectively. The brightest maser feature in the burst epochs (after 2015.1) is shown in Figure \ref{fig:reference_feature}. The intensity, velocity and FWHM ranges of all the maser features were highly variable between epochs. In 2014.9 the brightest maser feature had an amplitude of 55 Jy beam$^{-1}$, while in 2015.3 the feature flared to 729 Jy beam$^{-1}$. Between the features in all the epochs, the centre velocity $V_\text{centre}$ ranged from $-49$ \kms{} to $0.92$ \kms{}. The FWHMs of single features ranged from 0.62 \kms{} to 2.5 \kms{}. Some maser clusters had a double Gaussian spectral distribution.
\cpar{}
In CM2-W1, only $1-3$ features with \vlsr{} $\sim -8$ \kms{} were detected in the early epochs, before and at the onset of the accretion burst. In CM2-W2, the masers were initially only features in the range of $-5 $ \kms{} $ \leq V_\text{LSR} \leq -12.5$ \kms{}, and the size of the bow shock traced by the water masers was $\sim 120$ AU. In the later epochs, highly blue-shifted features with \vlsr{} $\sim -50$ \kms{} were detected at the northern part of the bow structure. The masers also traced a more well-defined bow shape with a size of $\sim 420$ AU in later epochs. This is seen in the top of Figure \ref{fig:VERA_spotmaps}, where the maser spots become a straight line over time. CM2-W2 was the region containing the brightest masers, and also the masers which had the most significant flares due to the accretion burst \cite{2018MNRAS.478.1077M}. In the epochs 2014.7 $-$ 2015.3,  the masers in MM1-W1  form a linear structure on a milliarcsecond scale with a size of $\sim 15$ AU and velocity range $-4$ \kms{} $\leq V_\text{LSR} \leq 0$ \kms{}. After epoch 2015.3, the masers in MM1-W1 are displaced and the maser spots have a smaller linear structure, with size $\sim 7$ AU. The velocity range stayed the same. 
\cpar{}
The masers in the southern regions near the UCHII region NGC6334F displayed variability. UCHII-W1, the southernmost maser association, consists of multiple subclusters distributed over 600 AU with a velocity range spanning $-37$ \kms{} $\leq V_\text{LSR} \leq -5$ \kms{}. Most of the clusters consisted of single maser features. The clusters show normal proper motion in the epochs 2014.7 $-$ 2014.9. At 2015.1, one of the clusters at $-33.7$ \kms{} disappears. Between 2016.1 and 2016.2, three clusters disappeared, and only three clusters were detected. The brightest cluster, with a constant radial velocity range $-16$ \kms{} $\leq V_\text{LSR} \leq -8$ \kms{} was detected in all epochs. There were some morphological changes in this cluster, with maser features excited and disappearing between 2015.1 and 2016.2. UCHII-W2 consisted of a single feature at $-30$ \kms{} between 2014.7 and 2015.1, which was not detected in subsequent epochs. In 2016.1 and 2016.2 a new feature was detected which showed a double-peaked spectral profile in the velocity range $-31$ \kms{} $\leq V_\text{LSR} \leq -26$ \kms{}. UCHII-W3 had no detections before 2015.1, with detections of a $-31$ \kms{} feature in 2015.1 and 2015.3. This feature was not detected in 2015.9 and two new features, with \vlsr{} $-36$ \kms{} and $-49$ \kms{} were detected in the final two epochs. 
\cpar{}
Figure \ref{fig:reference_feature} shows the spot positions and spectral profile of the single maser feature throughout all the epochs. This feature was in the bow shock CM2-W2. The spatial distributions of the pre-burst maser spots are compact with a size of $470$ $\mu$as corresponding to a linear scale of 0.6 AU. After the burst, the feature subtended an angle of $1060$  $\mu$as, corresponding to a linear scale of 1.4 AU. The feature's spectral profile also shows significant variability. Table \ref{tab:reference_feature} shows the Gaussian parameters for the feature. The feature changed from a single to a double Gaussian between 2014.7 and 2014.9, and back to a single Gaussian in 2015.3. The feature intensity was also variable and was the brightest feature in the field for all epochs. HartRAO monitoring observations have shown a high cadence time series which can be attributed to this feature \cite{2018MNRAS.478.1077M}. The time series shows that the maser flux density sharply rose during the onset of the accretion burst, then dimmed slightly, after which the flux density was relatively constant. We see the same behaviour in our observations. 
\begin{table}[h!]
    \centering
    \begin{tabular}{rrrr}
       Epoch & $A$ (Jy beam$^{-1}$) & $V_\text{centre}$ (km s$^{-1}$) & FWHM (km s$^{-1}$) \\
       \hline \hline
    2014.7 &  112 (2) & $-$7.26 (0.02) & 1.84 (0.04) \\
   2014.9 &   41 (1) & $-$7.52 (0.04) & 2.24 (0.08) \\
    &   29 (2) & $-$5.95 (0.03) & 1.30 (0.05) \\
   2015.1 &  163 (6) & $-$8.38 (0.09) & 2.05 (0.16) \\
    & 172 (10) & $-$6.53 (0.07) & 1.70 (0.10) \\
    2015.3 & 755 (19) & $-$7.18 (0.03) & 2.17 (0.06) \\
    2015.9 &  351 (7) & $-$7.61 (0.01) & 1.46 (0.03) \\
    2016.1 & 465 (38) & $-$7.61 (0.05) & 1.32 (0.12) \\
    2016.2 & 483 (19) & $-$7.55 (0.03) & 1.46 (0.07) \\
    \end{tabular}
    \caption{Gaussian fit parameters for the  maser feature shown in Figure \ref{fig:reference_feature}. The values in parenthesis are the uncertainties on the fit. Epochs 2014.9 and 2015.1 had a double Gaussian feature, and the parameters of both Gaussians are shown in this table.}
    \label{tab:reference_feature}
\end{table}
\section{Discussion}
\subsection{Large and small scale effects of the accretion burst on the water masers}
There were changes both in the spatial and velocity distribution of the water masers on a large ($10 - 100$ AU) and small ($0.5 - 5$ AU) scale. The masers in CM2 flared significantly, and appear to trace the bow-shaped shock front more closely. In the southern regions, some associations disappeared, and new features were activated \cite{2021saip.confE...1V}. In all epochs, the large-scale bipolar spatial distribution was visible.
\cpar{}
The $-7.6$ \kms{} feature shown in Figure \ref{fig:reference_feature} is the brightest in the field in the epochs after 2015.3. The interval between 2015.1 and 2015.3 was a time when the maser flared rapidly \cite{2018MNRAS.478.1077M}. The change in the linear size of the feature possibly indicates that an increased volume of water molecules had a pumping rate able to sustain maser action. There are some possible explanations. Water masers are known to support population inversion by collisions with H$_2$ and emission at specific ``sink" transitions \cite{2013ApJ...773...70H}. The accretion burst could amplify the collisional pump either mechanically or radiatively. CM2-W2 is $\sim 3000$ AU from the accretion bursting source MM1B, with an excavated cavity along the jet axis \cite{2018ApJ...866...87B}. Material ejected by the burst with a speed of 150 \kms{} would take 95 years to reach CM2-W2, ruling out mechanical amplification \cite{2021ApJ...908..175C}. With regards to radiative amplification, a stronger radiation field could cause the maser flare if the collisional partner has a high opacity to the radiation, and if a portion of the energy from the radiation is converted to kinetic energy in the collisional partner H$_2$ \cite{1978ApJ...226L..21B}. At high densities ($n \geq 10^6$ cm$^{-3}$) H$_2$ can be thermally heated by high-energy radiation \cite{1989ApJ...338..197S}. It is possible that high-energy radiation propagated freely through the excavated cavity, heating up the H$_2$ which would in turn increase the pump rate. Our proposal builds on the proposal of \cite{2018ApJ...866...87B}.
\cpar{}
Recently, a radiative pumping scheme for 22 GHz water masers was proposed \cite{2022MNRAS.513.1354G}. They showed that population inversion can be sustained radiatively for dust temperatures $\sim 1400$ K, water number densities $10^{4-6}$ cm$^{-3}$ and hydrogen kinetic temperatures $< 500$ K. The density and kinetic temperature requirements are reasonable if you take the shock model of \cite{2013ApJ...773...70H} into account, but it would have to be shown that CM2-W2 is under the effect of a 1400 K dust temperature radiation field for this explanation to be valid.
\subsection{Identifying maser features and the spectral method of proper motion calculation}
In water maser proper motion measurements, an important step after calibration and imaging is identifying the persistent object that you are tracking over time. There are two main approaches to identifying persistent water masers. These approaches can be called the ``spot" and the ``spectral" method.
\cpar{}
The spot method consists of using individual maser spots in a single velocity channel as a persistent object. The foundational assumption of the spot method is that the  velocity drifts are smaller than the instrument's channel width throughout the observations. This approach was used in proper motion calculations by e.g. \cite{2021ApJ...908..175C}. The spectral method involves using the intensity-weighted centroid position of the three brightest spots in a maser spectral feature. The feature method assumes that a single Gaussian spectral profile indicates a single masing cloud. This method has been used by e.g. \cite{2017MNRAS.467.2367B}. 
\cpar{}
We argue that the feature method is the most reliable and robust method of calculating proper motions for masers. Some data sets have calibration errors leading to spurious detections due to side lobes. The spot method has no built-in way to discern these spurious detections. On the other hand, the spectral method requires a Gaussian spectral distribution, which is not found for side-lobe detections. Further, the spectral method does make an assumption about the velocity drift and can take velocity drift into account if the feature can be identified to be persistent. The requirement for the spectral method is that the spectral resolution of the observations must be much smaller than the line width. For observations with a coarse spectral resolution, the spectral and spot methods are equivalent, as no Gaussian will be seen in the spectral profile. 
\cpar{}
In summary, calculating the proper motion with the intensity-weighted centroid position of multiple maser spots which are spatially close on the order of 1 AU and which show a single Gaussian in its spectral profile, is a more robust and reliable way to calculate proper motions than by using spots in a single channel.
\section{Conclusion}
We report multi-epoch VLBI observations of 22 GHz water masers with VERA before and during the recent accretion burst in NGC6334I. We identified around 20 clusters of individual maser spots with a Gaussian spectral distribution and calculated their amplitudes, centre velocities and FWHMs. We found all these parameters to be variable between pre-burst and burst epochs in all regions where we detected water masers. The brightest maser feature in CM2-W2 was found to have a variable linear scale, with a pre-burst scale of 0.6 AU and a burst scale of 1.4 AU.  This possibly indicates that a larger volume of gas had the correct pumping conditions to sustain masing action. We proposed that the water maser flare could be explained by the amplification of the collisional pump rate through radiative heating of H$_2$ by increased high-energy radiation (UV or X-ray). Lastly, we also argued that proper motions are more reliably calculated with the intensity-weighted centroid of the brightest maser spots in a spectral feature, rather than using single spots. 
\section*{Acknowledgements}
JMV acknowledges the National Research Foundation of South Africa for funding this research (Grant Number: 134192). JOC acknowledges support from the Italian Ministry of Foreign Affairs and International Cooperation
(MAECI Grant Number ZA18GR02) and the South
African Department of Science and Technology’s National Research Foundation (DST-NRF Grant Number
113121) as part of the ISARP RADIOSKY2020 Joint
Research Scheme. T. Hirota
is financially supported by the MEXT/JSPS KAKENHI Grant Number 17K05398.  This research made use of NASA’s
Astrophysics Data System Bibliographic Services. VERA is operated by Mizusawa VLBI Observatory, a branch
of the National Astronomical Observatory of Japan. 
\section*{References}
\bibliographystyle{iopart-num}
\bibliography{iopart-num.bib}

\end{document}